\documentclass[preprint2]{aastex}

\shorttitle{OO Aql}
\shortauthors{{\.I}\c{c}li et al.}

\begin{document}

\title{The low-mass interacting binary system OO Aql revisited: a new quadruple system}
\author{T. {\.I}\c{c}li\altaffilmark{1}, D. Ko\c{c}ak\altaffilmark{1},  G. \c{C}. Boz\altaffilmark{1} and K. Yakut\altaffilmark{1,2}}
\affil{Department of Astronomy and Space Sciences, University of Ege, 35100, Bornova--{\.I}zmir, Turkey}
\affil{Institute of Astronomy, University of Cambridge, Madingley Road, Cambridge CB3 0HA, UK}

\begin{abstract}
In this study we present photometric and spectroscopic variation analysis and orbital period study
of a low-mass interacting system OO Aql. Simultaneous solution of the light and radial velocity curves  provide us a determination of  new
set of stellar physical parameters for the primary and the secondary companion as
M$_{1}$ = 1.05(2) M$_{\odot}$, M$_{2}$ = 0.89(2) M$_{\odot}$, R$_{1}$ = 1.38(2) R$_{\odot}$,
R$_{2}$ = 1.28(2) R$_{\odot}$, $\log{(L_1/L_{\odot})} = 0.258$ and $\log{(L_2/L_{\odot})}  = 0.117$
and the separation of the components were determined a = 3.333(16) R$_{\odot}$.
Newly obtained parameters yield the distance of the system as 136(8) pc.
Analyses of the mid-eclipse times indicate a period increase of $\frac{P}{\dot{P}}=4\times 10^{7}$ yr
that can be interpreted in terms of the  mass transfer $\frac{dM}{dt}=5\times 10^{-8}$ M$_{\odot}$/yr
from the less massive component to the more massive component.
Our new solution confirmed that OO Aql is a multiple system in the form of AB + C + D.
We found initial astrophysical parameters for the component of the system and its
current age to be 8.6 Gyr using nonconservative stellar evolution model (EV-TWIN code).
\end{abstract}

\keywords{Stars: binaries -stars: binaries: close - stars: individual: OO Aql - stars: fundamental parameters
-stars: late-type - stars: evolution}

\section{Introduction}
One of the fundamental problems of modern astrophysics in stellar structure and evolution is the evolution of interacting binary systems.
The evolution of stellar components in close binary systems depend on several different physical processes than the solar like stars evolution depend.
Mass loss due to the stellar activity as a result of escaping plasma from the star, mass transfer between
the stellar components and angular momentum loss are the most crucial parameters in the evolution of interacting systems.
In addition, a third (or fourth) component orbiting the system will also affect the orbital period and therefore the
 evolution of a binary system. One way to get detailed information on stars is to determine their orbital and physical parameters.
 We selected the multiple system OO Aql for our project. Its physical parameters are obtained precisely
 and therefore it is quite a convenient system to examine a multiple interacting system's evolutionary stages.

G-type contact binary system OO Aql (HD 187183, $\alpha_{2000}=19~48~12.65$, $\delta_{2000}=+09~18~32.38$, V=$9\fm49$, B-V=$0\fm77$)
was classified as a variable star by Hoffleit (1932). Than it has been observed by Binnendijk (1968), Pohl (1969),
Pohl \& K{\i}z{\i}l{\i}rmak (1970, 1975), Djurasevic \& Erkapic (1998) and by Lafta \& Grainger (1985).
Djurasevic \& Erkapic (1998) studied the system's light variation in detail and analysed the light curves
obtained in different observing seasons. Analysis of the system yielded the Roche lobe filling factor ($f$)
to be 0.08  and the orbital inclination angle to be $86^{\circ}$. At the same study, the authors were also provided
a detailed model of a stellar spot.

OO Aql has an unusually high mass ratio. Spectroscopic studies of the OO Aql were provided by Hrivnak (1989), Hrivnak et al. (2001) and Pribula et al. (2007).
Hrivnak (1989) studying the Ca and H lines derived the semi-amplitude of the radial velocity curve and the mass function of the components.
He obtained a mass ratio of 0.843(8). Hrivnak et al. (2001) studied IUE satellite observations of the system and they reported a variation of Mg II \textit{h} and \textit{k} lines.
By using photometric and spectroscopic data Hrivnak et al. (2001) derived the mass of the components to be
M$_{1}$ = 1.05 (2) M$_{\odot}$  and  M$_{2}$ = 0.88 (2) M$_{\odot}$ and the radius of the components to be R$_{1}$ = 1.38 (2) R$_{\odot}$ and  R$_{2}$ = 1.28 (2) R$_{\odot}$ .
Recently, accurate radial velocities were obtained by Pribula et al. (2007).
They derived the semi-amplitude of the radial velocity curve and the mass function of the components
to be K$_1=153.03$ km\,s$^{-1}$, K$_2=180.81$ km\,s$^{-1}$, (M$_1$+M$_2$)$\sin^3 i = 1.954(19)$  M$_{\odot}$.

The orbital period variation analysis of OO Aql have been done by Binnedijk (1968),
Essam (1992) and Borkovits et al. (2005). Essam and Binnedijk's study reveal a parabolic
variation in other words the existence of a mass transfer. Following study by Demircan \& G{\"u}rol (1996)
discusses the parabolic and sinusoidal variation. The authors  detected the presence of a third body
orbiting the binary system with a period of 89 years. Borkovits et al. (2005) analyses reveal a sinusoidal
variation as a result of the third body orbiting the system with about 75 years orbital period in an
eccentric orbit of 0.06. The calculated mimimum mass of the thirtary component is 0.7 M$_{\odot}$. Finally, the results
obtained by Zache (2005) show parabolic and sinusoidal variations due to the presence of  a third body
with 72 years period and discuss the existence of a fourth body.

In this study newly obtained $VRI$ colours light curves and published radial velocity curves are analysed simultaneously.
The orbital and physical parameters of the system as well as the distance of the system derived precisely.
Following the observational information given in the second section, period variation study presented
in the third section. Light and radial velocity models are presented in section four and
in the fifth section physical parameters of the binary system are presented.
Finally, the evolution of the binary system is discussed and the results are summarized.

\section{New Observations}
\begin{figure}
\includegraphics[width=80mm]{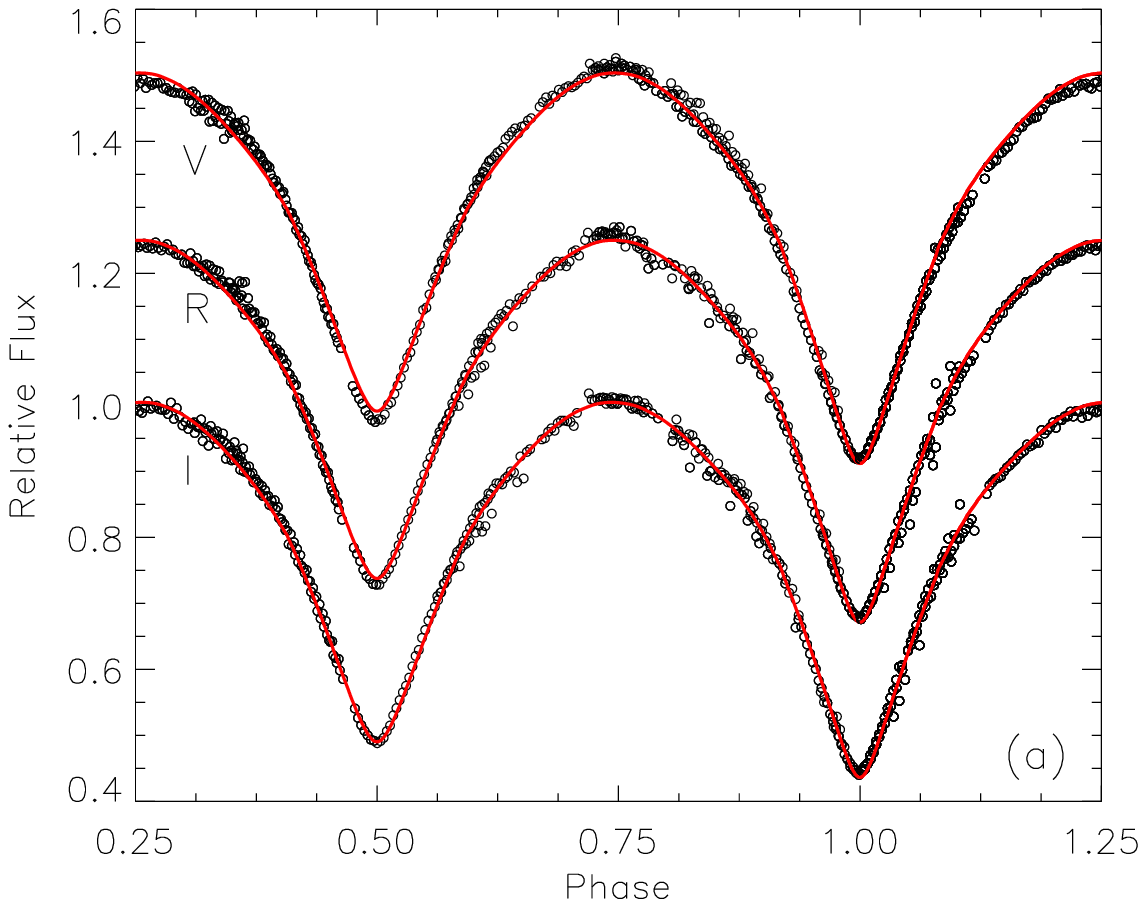}\\
\includegraphics[width=80mm]{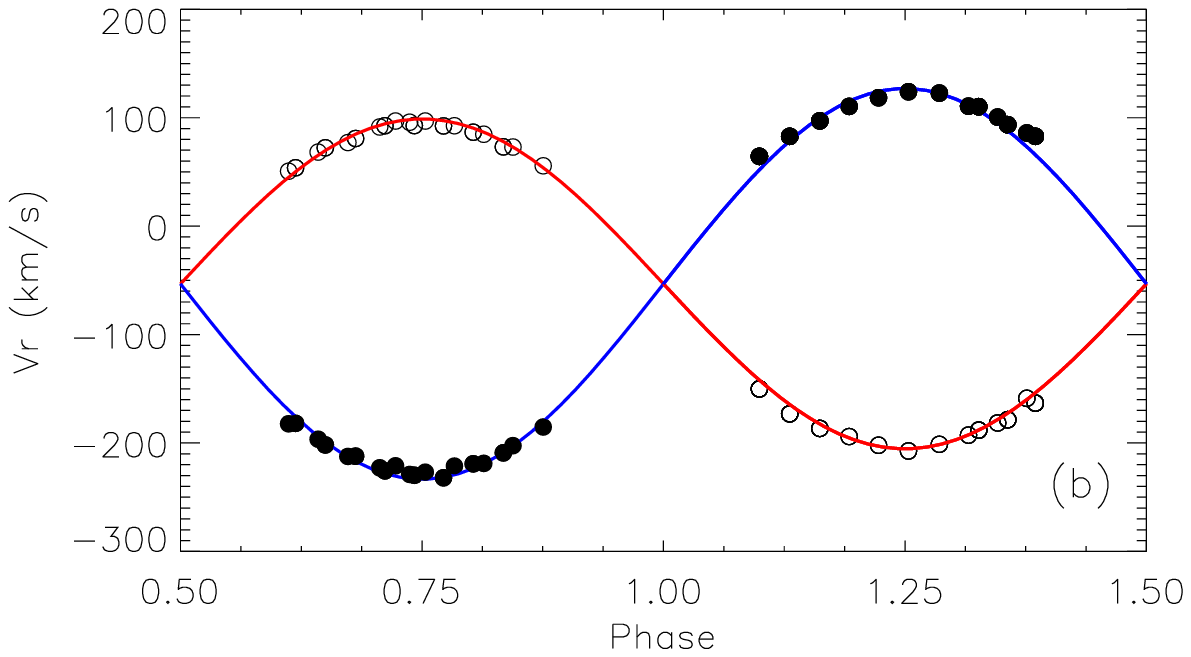}
\caption{(a) V, R and I observations of OO Aql obtained at EUO. TYC 1058-689-1 was used as a comparison star.
The solid line shows the theoretical fit obtained using the parameters given in Table~\ref{tab:OOAql:lcanalysis}.
(b) The radial velocity variation of OO Aql system. The observational points were obtained from Pribula et al.'s study.}\label{Fig:OOAql:LC}
\end{figure}

New light variations of the system OO Aql were obtained using 40cm diameter telescope at the
Ege University Observatory (EUO) at 10 nights. $VRI$ filters were used. Observations were performed on 20, 21, 22, 27, 28, 29 July;
4, 5 August and on 8, 9 September 2012. On 8, 9 September only times of minima were obtained.
Apogee CCD 2048X2048 was used  during the observations. TYC 1058-689-1 and TYC 1058-409-1 were
selected as comparison stars as they have been used in previous studies presented in the  literature.
Each nights observations are reduced separately with IRAF/APPHOT packaging.
During reduction processes each nightly bias correction and then dark are removed from all the
images and divided by flat images to obtain scientific frames. Differential photometry of these
calibrated images were used to obtain the difference in magnitudes (V-C).

Observation errors for V, R and I filters were 0.011, 0.007 and 0.007, respectively.
In Fig.~\ref{Fig:OOAql:LC} we show the light variations (V-C) in VRI colors. As shown in Fig.~\ref{Fig:OOAql:LC},
there is a difference in maxima which is not as prominent as obtained by
Djurasevic \& Erkapic (1998). Discussion on this issue will be re-addressed in the Section 4.
We obtained 3 new minima times throughout these new observations.
They are collated with those published and listed in Table~\ref{tab:OOAql:mintimes} with their errors.

\section{Period Variation Analyses}
There are many factors that may change the orbital period of a close/interacting binary system.
The most important of these factors are mass transfer between the stellar component stars,
mass loss due to stellar activity and the third (or fourth) bodies dynamic effects.
Especially on close interacting binary systems these effects can be identified remarkably by observations.
We can estimate this variation by measuring accurate times of minima light.
Times of minima observations that spread over many years allow us to determine the nature of the variation better.
Systems such as OO Aql are good candidates to find out these types of variations.
The difference between the observed (O) and calculated (C) times of minima in an eclipsing system
can provide us information about any orbital period variation(s).

A stellar component of a contact binary system transfers mass to its companion from L1 point.
This causes an increase/a decrease in orbital period. This change depends
on the mass of the mass transferring and accreting stars. In addition, in presence of a
third body periodic orbital period variation occurs. In the case of mass transfer
as well as a in existence of a third body the O-C variations show a sinusoidal variation superimposed on
a parabolic variation. OO Aql is a contact system where one of the component losses
 mass to its companion with a third/fourth component orbiting the binary system.
 Hence, both parabolic and double-sine-like variation are expected in its O-C variation.

 \begin{equation}
\begin{array}{l}
\textrm{Min\,I} = T_o +P_oE + \frac{1}{2} \frac{dP}{dE}E^2  \\
\\
+ \frac {a_{12} \sin
i'}{c}  \left[ \frac {1-{e'}^2}{1+{e'} \cos v'} \sin \left( v' +
\omega' \right) +{e'} \sin \omega' \right]
 \label{Eq:OOAql:1}
\end{array}
\end{equation}

These kind of changes are usually expressed in the form given in Eq.(\ref{Eq:OOAql:1}).
$T_{\rm o}$ and $E$ are the starting epoch for the primary minimum and eclipse cycle number, $P_{\rm o}$ is the orbital period of the
binary $a_{12}$, $i'$, $e'$ and $\omega'$ are the semi-major axis, inclination, eccentricity and the
longitude of the periastron of eclipsing pair about the third body
and $v'$ denotes the true anomaly of the position of the center of
mass (\emph{see} Kalomeni et al. 2007 for details).
The first two terms of the equation are linear while the third term represents a parabolic
variation due to the mass transfer and the fourth term represents the effect of a third
body.

\begin{table*}
\begin{center}
\tiny
\caption{Times of minimum light for OO Aql.}\label{tab:OOAql:mintimes}
\begin{tabular}{llllllllllllll}
\hline
HJD Min     &   Ref &   HJD Min     &   Ref &   HJD Min     &   Ref &   HJD Min     &   Ref &   HJD Min     &   Ref &   HJD Min     &   Ref &   HJD Min     &   Ref  \\
+2400000    &       & +2400000      &       &   +2400000    &       &   +2400000    &       &   +2400000    &       &  +2400000     &       &  +2400000      &   \\
\hline
33925.3742	&	1	&	41571.3468	&	7	&	47060.3709	&	19	&	48853.3865	&	23	&	51679.5093	&	31	&	52842.3471	&	44	&	53966.4152	&	50	\\
34194.4820	&	1	&	41571.3476	&	7	&	47061.3846	&	20	&	48854.3987	&	23	&	51780.3608	&	34	&	52846.4022	&	37	&	53967.4290	&	50	\\
34518.8310	&	2	&	41626.3310	&	7	&	47061.3853	&	20	&	48854.3994	&	23	&	52107.4975	&	35	&	52904.1768	&	47	&	53968.4424	&	50	\\
34600.4255	&	1	&	41890.3721	&	7	&	47263.5923	&	17	&	48854.3996	&	23	&	52121.4337	&	36	&	53173.5372	&	48	&	53995.3022	&	56	\\
37200.5320	&	3	&	41922.2987	&	7	&	47326.4345	&	17	&	49179.5059	&	23	&	52137.3988	&	34	&	53184.4335	&	49	&	54000.3692	&	50	\\
37879.6377	&	4	&	41940.2886	&	7	&	47326.4348	&	17	&	49179.5068	&	23	&	52175.4075	&	34	&	53218.3895	&	49	&	54335.3602	&	37	\\
37915.6215	&	4	&	42247.4023	&	8	&	47380.4049	&	18	&	49186.3467	&	23	&	52440.4595	&	37	&	53220.4148	&	49	&	54335.3602	&	37	\\
37929.5597	&	4	&	42252.4697	&	8	&	47380.4049	&	18	&	49186.3471	&	23	&	52481.5095	&	38	&	53226.4964	&	50	&	54335.3604	&	37	\\
37932.6007	&	4	&	42252.4704	&	8	&	47380.4049	&	18	&	49186.3473	&	23	&	52497.4739	&	39	&	53253.3553	&	37	&	54596.8635	&	57	\\
38239.4660	&	3	&	42301.3758	&	9	&	47381.4181	&	18	&	49186.3475	&	23	&	52498.4877	&	39	&	53258.4244	&	49	&	54639.6886	&	57	\\
38641.6028	&	4	&	42311.2580	&	9	&	47464.2828	&	18	&	49193.4417	&	23	&	52505.3277	&	40	&	53258.4244	&	49	&	54668.8275	&	57	\\
38645.6576	&	4	&	42960.4541	&	10	&	47803.3205	&	18	&	49193.4419	&	23	&	52508.3695	&	40	&	53530.5719	&	48	&	54675.4163	&	58	\\
39300.6810	&	3	&	42986.8086	&	11	&	47805.3476	&	21	&	49193.4423	&	23	&	52514.4484	&	41	&	53544.5103	&	51	&	54675.4165	&	58	\\
39322.7288	&	4	&	43370.4473	&	12	&	48109.4264	&	22	&	49193.4429	&	23	&	52514.4503	&	41	&	53569.3419	&	51	&	54681.7517	&	57	\\
39327.7967	&	4	&	44460.8016	&	13	&	48475.3217	&	23	&	49238.2925	&	23	&	52514.4508	&	41	&	53588.3462	&	51	&	54684.5386	&	59	\\
39341.7338	&	4	&	44476.7681	&	13	&	48477.3477	&	23	&	49544.3920	&	24	&	52548.4080	&	42	&	53591.3870	&	49	&	54708.3580	&	60	\\
39714.7280	&	3	&	45180.4450	&	14	&	48477.3479	&	23	&	49550.4800	&	25	&	52574.2521	&	39	&	53606.5916	&	52	&	54727.6159	&	60	\\
40068.4638	&	3	&	46705.3655	&	15	&	48500.4081	&	21	&	49924.4856	&	26	&	52575.2659	&	39	&	53607.6051	&	53	&	54730.4054	&	60	\\
40366.4540	&	3	&	46980.2979	&	16	&	48530.3088	&	23	&	50363.3710	&	27	&	52576.2796	&	42	&	53609.3790	&	48	&	54976.4513	&	59	\\
40811.4152	&	5	&	46981.3123	&	16	&	48530.3099	&	23	&	50714.3240	&	27	&	52780.5145	&	43	&	53613.4327	&	55	&	54978.4790	&	50	\\
40817.4940	&	5	&	46981.3131	&	16	&	48531.3224	&	23	&	50718.3750	&	27	&	52781.5316	&	43	&	53615.7139	&	53	&	55018.5154	&	61	\\
40825.3515	&	5	&	46976.4984	&	17	&	48532.3359	&	23	&	50719.3922	&	28	&	52801.5506	&	44	&	53622.5555	&	53	&	55101.3767	&	62	\\
40858.2910	&	5	&	47024.3908	&	18	&	48838.4357	&	23	&	50967.4608	&	29	&	52811.4324	&	43	&	53881.5279	&	55	&	55385.4336	&	60	\\
41161.3518	&	6	&	47028.4462	&	18	&	48838.4360	&	23	&	51040.4440	&	30	&	52813.4596	&	45	&	53934.4883	&	37	&	55487.2998	&	60	\\
41179.3457	&	6	&	47058.3422	&	19	&	48838.4368	&	23	&	51380.5004	&	31	&	52813.4599	&	46	&	53935.5008	&	37	&	56129.4061	&	63	\\
41182.3845	&	6	&	47059.3597	&	18	&	48853.3862	&	23	&	51393.4238	&	32	&	52815.4857	&	46	&	53963.3748	&	37	&	56130.4205	&	63	\\
41187.4531	&	6	&	47059.3625	&	18	&	48853.3863	&	23	&	51456.7720	&	33	&	52822.3298	&	44	&	53966.4150	&	55	&	56145.3701	&	63	\\
\hline
\end{tabular}
\end{center}
\scriptsize {References for Table~\ref{tab:OOAql:mintimes}.
1-Kwee (1958); 2- Fitch (1964); 3-Pohl et. al.  (1969); 4-Binnendijk (1968); 5-K{\i}z{\i}l{\i}rmak \& Pohl (1971); 6-Pohl \& K{\i}z{\i}l{\i}rmak (1972)
;7-K{\i}z{\i}l{\i}rmak \& Pohl (1974); 8- Pohl \& K{\i}z{\i}l{\i}rmak (1975); 9-Pohl \& K{\i}z{\i}l{\i}rmak (1976); 10-Pohl \& K{\i}z{\i}l{\i}rmak (1977);
11-Scarfe (1978); 12- Ebersberger et al. (1978); 13-Scarfe et al.(1984); 14-Pohl et al.(1983); 15-Pohl et al.(1987); 16-Essam et. al. (1992); 17-Keskin \& Pohl(1989);
18-Hubscher \& Lichtenknecker (1988); 19-Hegedus (1987); 20-Hanzl(1990); 21-Wunder et al.(1992); 22-Hanzl (1991); 23-G{\"u}rol (1994); 24-Diethelm (1994); 25-Acerbi (1994);
26-Agerer \& Hubscher (1996); 27-Agerer \& Hubscher (2000); 28-Agerer \& Hubscher (1998); 29-Borkovits \& Barna (1998); 30-Agerer \& Hubscher (1999);
31-Biro \&  Borkovits (2000); 32-Agerer \& Hubscher (2001); 33-Nelson (2000); 34-Agerer (2002); 35-Borkovits et al. (2001); 36-Agerer \& Hubscher (2003); 37-Brat et al. (2007);
38-Borkovits et al. (2002); 39-Selam et al. (2003); 40-Demircan et al. (2003); 41-G{\"u}rol et. al.(2003); 42-Diethelm (2003); 43-Bak{\i}{\c s} et. al. (2003);
44-Bak{\i}{\c s} et. al. (2005); 45-Biro et. al. (2006); 46-Hubscher (2005); 47-Kim et. al. (2006); 48-Hubscher et. al. (2006); 49-Marino et al. (2010); 50-Hubscher et. al.(2005);
51-Senavci et al. (2007); 52-Parimucha et al. (2009); 53-Alton (2006); 54-Biro et.al.(2007); 55-Hubscher \& Walter (2007); 56-Do{\u g}ru et al. (2007);
57-Samolyk (2008); 58-Brat et al. (2008); 59-Hubscher et.al.(2009); 60-Hubscher (2011); 61-Erkan et al. (2010); 62-Borkovits et al. (2011); 63-This study }
\end{table*}

A total of 189 both primary and secondary times of minima light, obtained with photoelectric
and CCD techniques with three new minimum times presented in this study, were used in analysis.
All collected minimum times are listed in Table~\ref{tab:OOAql:mintimes}.
Those times of minima with  Eq.(\ref{Eq:OOAql:1}) were analysed by the least squares method
using the ephemeris $\textrm{HJD MinI} = 2438239.720 + 0.5067883 \times E$ given in Demircan \& G{\"u}rol (1996) as initial ephemeris.
During analysis the visual and photographic times of minima
showed so much scattering that they have been ignored. Instead, reliable photoelectric and
CCD observations were used in times of minima light analyses. All the times of minima lights
were given the same weights during the O-C analysis.

\begin{table}
\scriptsize
\caption{Orbital elements of the quadruple system OO Aql. The
standard errors 1$\sigma$, in the last digit are given in parentheses.} \label{tab:OOAql:OCResults}
\begin{tabular}{llllll}
\hline
Parameter                                       &Unit                          &  Value          \\
\hline
                                                &                             &                  \\
\textit{\textbf{Binary system -- Star AB}}      &                             &                   \\
Initial epoch, T$_{\rm o}$                      &HJD                          & 24~38239.696(2)   \\
Period, P$_\textrm{bin}$                        &day                          & 0.50679020(8)       \\
Period change ratio $\frac{P}{\dot{P}}$         &yr                           & $4(1)\times 10^{7}$ \\
Mass transfer ratio, $\frac{dM}{dt}$            &$\rm{M_{\odot}}$yr$^{-1}$    & $5(1)\times10^{-8}$   \\
Seperation between stars, $a$                   &$\rm{R_{\odot}}$             & 3.34(2)                \\
Parallax, $\pi$                                 &mas                          & 7.35                    \\
                                                                                &                          \\
\textit{\textbf{Star C}}                        &                               &                           \\
Initial epoch, T$_{\rm o}$(C)                   &HJD                          & 24~54392                  \\
Orbital period, $P_C$                           &yr                           & 20(1)                   \\
Amplitude, $A_C$                                &day                          & 0.0041(2)                 \\
Eccentricity, $e$                               &                             & 0.44(7)                         \\
Longitude of the periastron, $\omega'_C$        &$^\circ$                     & 164                       \\
Mass function, $f(m_C)$             l            &$\rm{M_{\odot}}$             & 0.0011                  \\
Minimum mass, M$_{C\textrm{(min)}}$             &$\rm{M_{\odot}}$             & 0.19                    \\
Mass, M$_{C;i'=60^\circ}$                       &$\rm{M_{\odot}}$             & 0.23                     \\
Mass, M$_{C;i'=30^\circ}$                       &$\rm{M_{\odot}}$             & 0.47                     \\
Angular distance, $a^"_C$                       &mas                          &76                       \\
                                                &                             &                             \\
\textit{\textbf{Star D}}                        &                             &                              \\
Initial epoch, T$_{\rm o}$(D)                   &HJD                          & 24~55622                    \\
Orbital period, $P_D$                           &yr                           & 52(2)                         \\
Amplitude, $A_D$                                &day                          & 0.019(1)                          \\
Eccentricity, $e$                               &                             & 0.220(20)                          \\
Longitude of the periastron, $\omega'_D$        &$^\circ$                     & 20                              \\
Mass function, $f(m_D)$                         &$\rm{M_{\odot}}$             & 0.0130                          \\
Minimum mass, M$_{D\textrm{(min)}}$             &$\rm{M_{\odot}}$             & 0.42                             \\
Mass, M$_{D;i'=60^\circ}$                       &$\rm{M_{\odot}}$             & 0.49                               \\
Mass, M$_{D;i'=30^\circ}$                       &$\rm{M_{\odot}}$             & 0.95                                \\
Angular distance, $a^"_D$                       &mas                          & 138                                \\
\hline
\end{tabular}
\end{table}

\begin{figure}
\includegraphics[width=90mm]{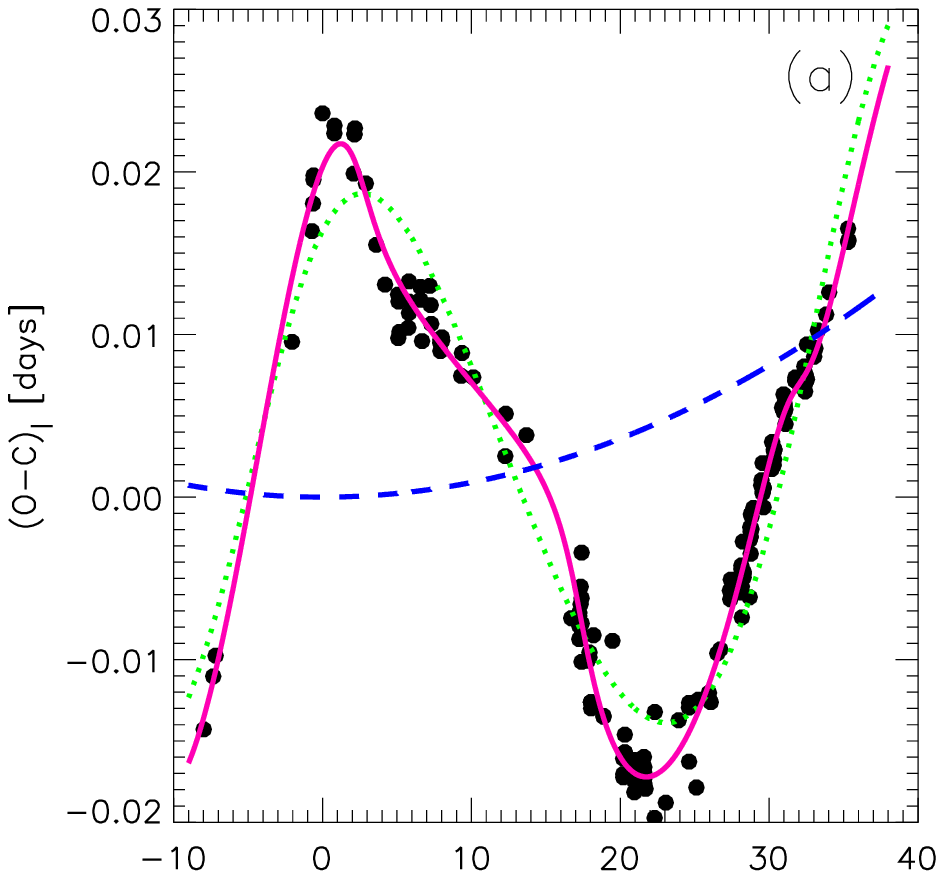}\vspace{-0.35in}\\
\includegraphics[width=90mm]{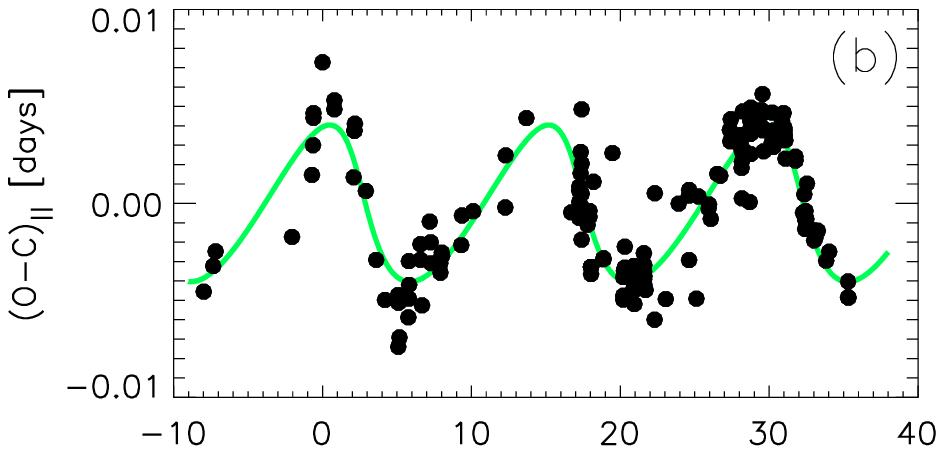}\vspace{-0.35in}\\
\includegraphics[width=90mm]{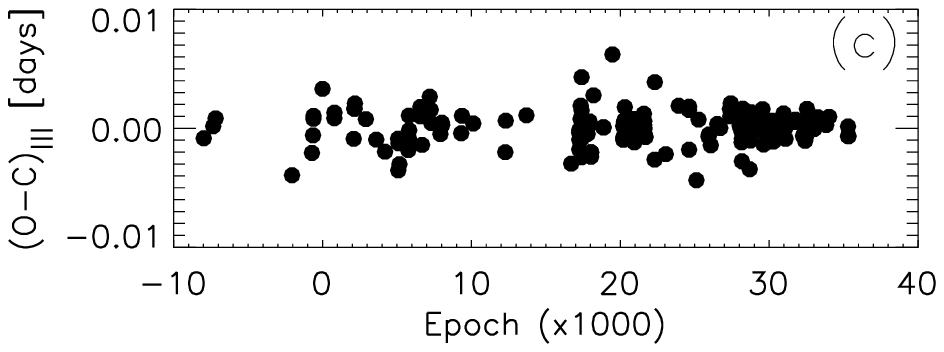}
\caption{(a) The O-C variation of OO Aql. The dotted line represents the O-C variation due to the third-body, while the
continuous line indicates the total effect of third and fourth stars. The dashed line shows the parabolic variation. (b) The (O-C)$_{\textrm{II}}$ residuals after the subtraction of parabolic change and third body orbit shown in (a). (c) shows the residuals.}\label{Fig:OOAql:OC}
\end{figure}

Using the Eq.(\ref{Eq:OOAql:1}) we analysed all the available minima time taking into account the mass transfer and a third body.
Residuals of the analysis show a sine-like variation (Fig.~\ref{Fig:OOAql:OC}b). The last term in the
Eq.(\ref{Eq:OOAql:1}),  therefore, is re-written for a fourth body and re-done the analysis. Fig.~\ref{Fig:OOAql:OC} shows the number of
cycles versus O-C variation. Fig.~\ref{Fig:OOAql:OC}a represents the effects of mass transfer and both
third and fourth bodies effects. Fig.~\ref{Fig:OOAql:OC}b shows only the fourth body effect and
Fig.~\ref{Fig:OOAql:OC}c show the residuals after removing each effects.

The parameters obtained from the orbital period analysis of OO Aql is listed in Table~\ref{tab:OOAql:OCResults}.
In this table the orbital elements for the third  body (Star D) and fourth body (Star C) are given separately.
The fourth body that is closer to the binary star is named as Star C and the furthest one is named as Star D.
Our analysis of OO Aql indicates the orbital period of the outer star as 52 years and the orbital period of the
interior body as 20 years. Our new solutions obtained for contact system OO Aql confirm that the system is a
multiple system in the form of AB + C + D.

\section{Simultaneous solutions of light and radial velocity curves}

V, R and I pass bands magnitudes are normalized and converted into normalized flux for light curve analyses.
During the normalization the mean magnitudes of V, R, I bands were taken as -0.976, -1.182 and -1.333, respectively.
Our new light variations with Pribula et al.'s (2007) radial velocity data were solved simultaneously.
Albedos ($A_{\rm 1}$, $A_{\rm 2}$) were obtained from Rucinski (1969) and gravity darkening coefficients
($g_{\rm 1}$, $g_{\rm 2}$) were taken from Lucy (1967). The logarithmic limb-darkening law were used with the
coefficients adopted from van Hamme (1993) for a solar composition star.
During solutions due to the uncertainties in observations different weights
were given for each colours. Solutions were done with Phoebe (Pr\~{s}a \& Zwitter 2005) that
is based on the Wilson-Devinney code (Wilson \& Devinney 1971).
The adjustable photometric parameters were orbital inclination, \textit{i}, surface potential, $\Omega_1 = \Omega_2 = \Omega$, temperature
 of the secondary component $T_{\rm 2}$, luminosity $L_{\rm 1}$ and the mass ratio $q$. The center of
 mass velocity $V_0$ and semi-major axis $a$  were also set as free parameters as well as the time of minimum light,
 $T_0$ and the orbital period, $P$. The obtained results are shown in Table~\ref{tab:OOAql:lcanalysis}.

The solution results are shown graphically with solid line in Fig.~\ref{Fig:OOAql:LC}a.
Open circles show observation points and the solid line presents theoretically obtained solution.
The radial velocity results obtained in the same solution are shown in Fig.~\ref{Fig:OOAql:LC}b.
The light curve analysis result is in agreement with observations. However, a
slight discrepancy near the phase of 0.65 in V color presents.
One reason for this discrepancy may be the stellar activity resulting as stellar spots.
O'Connell effect, the difference observed in maximum light, was not observed distinctively in this study.
For this reason, the light curve solution was done assuming no stellar spots.


\begin{table}
\scriptsize
\caption{Simultaneous analyses results of the light and radial velocity curve results and
their formal 1$\sigma$ errors for OO Aql. The indices $1$ and $2$ refers to the hot
and cooler components, respectively. See text for details.}
\begin{tabular}{lll}
\hline
Parameter                                   & Value      \\
\hline
$T_0$ (d)                                   & 24~49193.4990(12) \\
$P$ (d)                                     & 0.50678852(1) \\
$i$ ${({^\circ})}$                          & 85.6(1)   \\
$q = M_h / M_c$                             & 0.844(8)   \\
$a$ ($\rm{R_{\odot}}$)			            & 3.337(16)  \\
$V_0$ (kms$^{-1}$)			                & -53.3(7) \\
$\Omega _{1}=\Omega _{2}$                   & 3.391(3)      \\
$T_1$ (K)                                   & 5700  \\
$T_2$ (K)                                   & 5472(55)  \\
Fractional radius of primary comp.          & 0.4112(6) \\
Fractional radius of secondary comp.        & 0.3815(6) \\
$A_1 = A_2$                            & 0.6        \\
$g_1 = g_2$                            & 0.32        \\
Luminosity ratio:$\frac{L_1}{L_1+L_2}$ (\%)&   \\
$V$                                         & 58        \\
$R$   					                    & 58         \\
$I$   					                    & 57         \\
\hline
\end{tabular}
\label{tab:OOAql:lcanalysis}
\end{table}

\section{Physical Parameters}
The physical parameters of an eclipsing, double lined spectroscopic binary system OO Aql
can be obtained accurately. A sufficient number of the radial velocity data of the system
with the newly obtained accurate multi-color CCD observations allows us to determine system
parameters precisely. The effective
temperature of the Sun is taken to be 5777 K and its absolute magnitude was taken to be 4.732 mag while calculating the physical parameters of the component stars.
The mass of the primary star is obtained to be 1.05 $\rm{M_{\odot}}$  and the mass of its companion is obtained
to be 0.89 $\rm{M_{\odot}}$. The results obtained are slightly different from those exist in the literature.

The distance of the system is obtained to be 136(8) pc. This value is 13\% smaller than
the value given by SIMBAD.  The absorption effect of
the interstellar matter was ignored because the system is very close to us. Taking into account the distance of the system and the distance
of the third body and the fourth body to the binary system their angular distances were obtained
to be 0.138 arcsec and 0.074 arcsec, respectively. The Hubble Space Telescope's (HST)
resolving power is about 0.05 arcsec. This shows us that the 3rd and 4th body
can be seen as discrete sources in HST's images.

\begin{table}
\begin{center}
\scriptsize
\caption{Astrophysical parameters of the system. The standard errors
1$\sigma$ in the last digit are given in parentheses.}
\label{tab:OOAql:par}
\begin{tabular}{llll}
\hline
Parameter                                        &Unit                      & Primary           & Secondary   \\
\hline
Mass (M)                                         &$\rm{M_{\odot}}$        & $1.05(2)$            & $0.89(2)$      \\
Radius (R)                                       &$\rm{R_{\odot}}$        & $1.38(2)$            & $1.28(2)$      \\
Temperature ($\log T_{\rm eff}$)                 &$\rm{K}$                & 3.756(5)             &3.737(4)     \\
Luminosity (L)                                   &$L_{\odot}$             & $1.81(9)$            & $1.31(7)$      \\
Surface gravity ($\log g$)                       &$\rm{cms^{-2}} $        & $4.18$               & $4.17$      \\
Bolometric magnitude (M$_b$)                     &mag                     & 4.09                 & 4.44          \\
Absolute magnitude (M$_V$)                       &mag                     & 4.44                 & 4.73          \\

Distance (d)                                     &pc                      &~~~~~~~~136(8) &      \\
\hline
\end{tabular}
\end{center}
\end{table}

\section{Results and Conclusion}

In this study, OO Aql's VRI bands observations obtained at EUO
combined with Pribula et al.'s (2007) radial velocity observations  were solved simultaneously
and its physical and orbital parameters were obtained precisely.
The light and radial velocity observation results are given in Table~\ref{tab:OOAql:lcanalysis} and Table~\ref{tab:OOAql:par}.
The times of minima light spread over 61 years combined with those obtained in this study were analysed.
Analysis yield a mass transfer from the massive component to the less massive component
at a rate of $5\times10^{-8}$ $\rm{M_{\odot}}$  per year, an M star orbiting the system with an orbital period of 20
years with a solar like star with 52 years orbital period. Our solutions revealed that
the system is a quadruple system AB + C + D. In addition, estimations show HST can image this system discretely.

Observations of quadruple systems spread over many years are important to study the evolution
of quadruple systems as a laboratory. In this context OO Aql is an ideal laboratory to
study mass transfer and the angular momentum problem in the existence of a third and fourth body.
The majority of contact systems are generally thought to be multiple systems.
These systems are usually in the form of binary + star. On the other hand some of the
binary systems (e.g., XY Leo, Yakut et al. 2003) are in the form of  binary + binary.
Multiple systems like OO Aql binary+star+star known to be in a smaller number.

The evolution of close and interacting binary stars depend on the nuclear evolution, mass transfer,
mass loss and angular momentum loss. The effects of these processes are different on each stages of binary evolution.
A detached and close binary system during the early stages of main sequence evolve first with nuclear
evolution and dynamo based angular momentum loss. The evolution, then is driven by mass transfer of
Roche lobe filling primary star. Finally, the system evolves as a contact binary and mass and energy
transfer between the components continues. Evolution of close binary stars was discussed in details
by Yakut \& Eggleton (2005). The authors in that study discussed how contact binary systems
evolve and also defined a new energy transfer process.

\begin{figure}
\includegraphics[width=70mm]{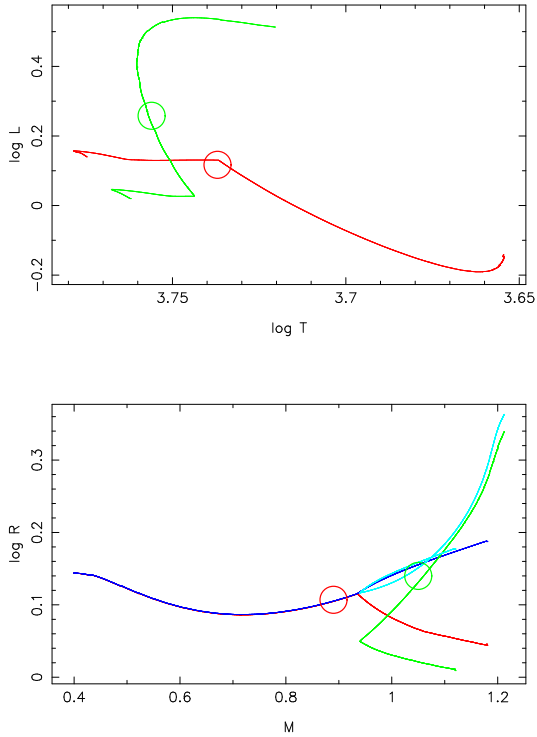}
\caption{Non-conservative evolution of OO Aql. More massive and less massive star are red and green; their respective
Roche-lobe radii are dark blue and light blue.
Initial parameters were 1.18 M$_{\odot}$, 1.12 M$_{\odot}$ and 0.63 days with Solar composition.
The original primary is the currently secondary component.}\label{fig:model}
\end{figure}

Using our newly obtained physical parameters we provide an evolutionary model of the
interacting binary OO Aql. We used the TWIN version of the EV code
(Eggleton, 1971, Pols et al. 1995, Eggleton \& Kiseleva-Eggleton 2002,
Yakut \& Eggleton 2005) that has been developed by Peter P. Eggleton.
We run dozens models using different initial parameters.
The best agreement with the observations was obtained for a model binary system with an initial
period of 0.63 days, with a primary and secondary masses 1.18 M$_\odot$ and 1.12 M$_\odot$.
Fig.~\ref{fig:model} shows the mass-radius diagram. Primary star's evolutionary track is
shown in red while its component is shown in green.  A system with an orbital
period of 0.63 days with 1.18$\rm{M_{\odot}}$  +1.12 $\rm{M_{\odot}}$  components that began to evolve at the
main-sequence system evolves 8.5 billion years later as a semi-detached binary
and a short time later it evolves as a contact system.
The system has reached today's mass and radius values at 8.7 billion years.

\section*{Acknowledgments}
We are very grateful to an anonymous referee for his/her comments and suggestions which helped us to improve the paper.
The authors would like to thank Peter P. Eggleton for his valuable comments and suggestions.
This study was supported by the Turkish Scientific and Research Council (T\"UB\.ITAK, Pr.2209 and 111T270).

\end{document}